\documentclass[aps,prc,twocolumn,superscriptaddress,showpacs]{revtex4}

\usepackage{graphicx}% Include figure files
\usepackage{longtable}
\usepackage{dcolumn}% Align table columns on decimal point
\usepackage{bm}% bold math
\usepackage{ulem}
\usepackage{color}
\usepackage{epsfig}

\begin{document}

\title{
Cross section measurement of the astrophysically important $^{17}$O(p,$\gamma$)$^{18}$F reaction in a wide energy range
}
\author{Gy. Gy\"urky}%
\email{gyurky@atomki.mta.hu}
\affiliation{Institute for Nuclear Research (Atomki), H-4001 Debrecen, Hungary}
\author{A. Ornelas}%
\affiliation{Institute for Nuclear Research (Atomki), H-4001 Debrecen, Hungary}
\affiliation{University of Debrecen, H-4001 Debrecen, Hungary}
\author{Zs. F\"ul\"op}%
\affiliation{Institute for Nuclear Research (Atomki), H-4001 Debrecen, Hungary}
\author{Z. Hal\'asz}%
\affiliation{Institute for Nuclear Research (Atomki), H-4001 Debrecen, Hungary}
\affiliation{University of Debrecen, H-4001 Debrecen, Hungary}
\author{G.G.~Kiss}%
\affiliation{Institute for Nuclear Research (Atomki), H-4001 Debrecen, Hungary}
\author{T. Sz\"ucs}
\affiliation{Institute for Nuclear Research (Atomki), H-4001 Debrecen, Hungary}
\author{R. Husz\'ank}
\affiliation{Institute for Nuclear Research (Atomki), H-4001 Debrecen, Hungary}
\author{I. Horny\'ak}
\affiliation{Institute for Nuclear Research (Atomki), H-4001 Debrecen, Hungary}
\author{I. Rajta}%
\affiliation{Institute for Nuclear Research (Atomki), H-4001 Debrecen, Hungary}
\author{I. Vajda}%
\affiliation{Institute for Nuclear Research (Atomki), H-4001 Debrecen, Hungary}

\date{\today}

\begin{abstract}
\begin{description}

\item[Background]

The $^{17}$O(p,$\gamma$)$^{18}$F reaction plays an important role in hydrogen burning processes in several different stages of stellar evolution. The rate of this reaction must therefore be known with high accuracy at the relevant temperatures in order to provide the necessary input for astrophysical models. 

\item[Purpose]

The cross section of $^{17}$O(p,$\gamma$)$^{18}$F is characterized by a complicated resonance structure at low energies which needs to be reproduced by theoretical models if a reliable extrapolation to astrophysical energies is required. Experimental data, however, is scarce in a wide energy range which increases the uncertainty of the extrapolations. The purpose of the present work is therefore to provide consistent and precise cross section values in a wide energy range for the $^{17}$O(p,$\gamma$)$^{18}$F reaction.

\item[Method]

The cross section is measured using the activation method. This method provides directly the total cross section which can be compared with model calculations. With this technique some typical systematic uncertainties encountered in in-beam $\gamma$-spectroscopy experiments can be avoided.

\item[Results]

The cross section was measured between 500 keV and 1.8 MeV proton energies with a total uncertainty of typically 10\,\%. The results are compared with earlier measurements and it is found that the gross features of the $^{17}$O(p,$\gamma$)$^{18}$F excitation function is relatively well reproduced by the present data. Deviation of roughly a factor of 1.5 is found in the case of the total cross section when compared with the only one high energy dataset. At the lowest measured energy our result is in agreement with two recent datasets within one standard deviation and deviates by roughly two standard deviations from a third one. An R-matrix analysis of the present and previous data strengthen the reliability of the extrapolated zero energy astrophysical S-factor. 

\item[Conclusions]

Using an independent experimental technique, the literature cross section data of $^{17}$O(p,$\gamma$)$^{18}$F is confirmed in the energy region of the resonances while lower direct capture cross section is recommended at higher energies. The present dataset provides a constraint for the theoretical cross sections.

\end{description}
\end{abstract}


\maketitle

\section{Introduction}
\label{sec:intro}

Hydrogen burning, the conversion of four protons into an alpha particle in the interior of stars, is the most important energy source in the universe and it is also responsible for the existence of several chemical elements. Besides the pp-chains powering e.g. our Sun, catalytic reactions cycles, like the various CNO cycles play the major role in hydrogen burning \cite{ade11,wie10}. Depending on the temperature and chemical composition of the stellar plasma, different CNO cycles can take place involving various isotopes of carbon, nitrogen, oxygen and fluorine. 

The $^{17}$O(p,$\gamma$)$^{18}$F reaction, which competes with the $\alpha$-emission in $^{17}$O(p,$\alpha$)$^{14}$N \cite{str16} is the starting point of the third CNO cycle. This cycle is activated in various stellar conditions such as red giant and asymptotic giant stars and classical novae. The abundances of fluorine and the heavy oxygen isotopes are strongly related to the operation of this cycle and therefore the rates of the participating reactions must be known.

Below 1.5\,MeV the $^{17}$O(p,$\gamma$)$^{18}$F reaction is characterized by many broad and narrow resonances. Therefore, the temperature dependence of the $^{17}$O(p,$\gamma$)$^{18}$F thermonuclear reaction rate shows a complicated picture. The rate depends on the direct capture component as well as on the narrow low energy resonances and the tails of the higher energy broad resonances (see e.g. Fig.\,10 in ref.\,\cite{dil14} for the contribution of the different components to the reaction rate). An R-matrix fit to the experimental data is therefore inevitable to provide reaction rates at various temperatures for stellar models.

The first cross section measurement of $^{17}$O(p,$\gamma$)$^{18}$F was carried out several decades ago by C. Rolfs in a wide energy range between 300\,keV and 1.9\,MeV \cite{rol73}. After the turn of the century, several experimental studies were carried out mostly concentrating on the low energy region below about 500\,keV \cite{fox04,fox05,cha05,cha07,new10,kon12,hag12,sco12,dil14,buc15}. (The only exception was the work of A. Kontos \textit{et al.} \cite{kon12} which extended up to 1.6\,MeV.) The comparison of the new precise data with the results of \cite{rol73} revealed some discrepancy both in the absolute scale and the energy dependence of the cross section at the lowest energies studied by \cite{rol73}.

In most of the previous experiments the cross section of $^{17}$O(p,$\gamma$)$^{18}$F was measured with in-beam $\gamma$-spectroscopy: the prompt $\gamma$-radiation from the formed $^{18}$F nucleus was detected. The complicated level scheme of $^{18}$F (see e.g. Fig.\,1. in ref.\,\cite{buc15}) implies that the detection of many primary and secondary transitions is necessary for the cross section determination. This represents a source of uncertainty in the experiments. In order to provide the astrophysically relevant total cross section, all the transitions must be measured and care must be taken to measure even the weakest $\gamma$-lines. The angular distributions of all the $\gamma$-emissions must also be known. Moreover, in order to measure low cross sections, close target-detector geometries are typically used leading to strong true coincidence summing effects.

All these experimental difficulties can be avoided by the application of the activation method, which was used by only two experiments before at energies below 400\,keV \cite{cha05,cha07,sco12,dil14}. The reaction product of $^{17}$O(p,$\gamma$)$^{18}$F is radioactive, decays by positron emission \footnote{A weak electron capture decay branching also exist with 3\,\% probability} with a half-life of 109.77\,$\pm$\,0.05\,minutes \cite{til95}. The decay is entirely to the ground state of $^{18}$O, no $\gamma$-radiation follows thus the decay. The emission of the 511\,keV $\gamma$-radiation following the positron annihilation, on the other hand, allows the measurement of the decay by $\gamma$-detection. By measuring the $^{18}$F activity the number of reaction product and therefore the total reaction cross section can be determined directly. The activation measurement of $^{17}$O(p,$\gamma$)$^{18}$F provides therefore an independent means of cross section determination which can be used to check earlier experimental data and provide a constraint for R-matrix calculations regarding the total reaction cross section.

The aim of the present work is therefore to measure the $^{17}$O(p,$\gamma$)$^{18}$F cross section with the activation method in a wide energy range. The next section provides detailed information about the experimental technique, the results are presented in Sec.\,\ref{sec:results} while Sec.\,\ref{sec:summary} provides the summary and conclusions.

\section{Experimental procedure}
\label{sec:experiment}

\subsection{Target preparation and characterization}
\label{subsec:target}

Solid state oxygen targets were produced by anodic oxidation of tantalum disks in water enriched in $^{17}$O. With this technique Ta$_2$O$_5$ layers can be produced with well defined Ta:O ratio and the targets have high stability under beam bombardment. The anodization setup was the same as used recently by the LUNA collaboration for the low energy $^{17}$O(p,$\gamma$)$^{18}$F cross section measurements \cite{sco12,dil14}. Full details of the anodization device and the preparation procedure have been published by the LUNA collaboration \cite{cac12}, here only the most important features and the differences are summarized. 

Two water samples were used for the target preparations. The isotopic abundances of the $^{16}$O, $^{17}$O and $^{18}$O isotopes, respectively, were the following: (15.5\,$\pm$\,0.6)\%, (77.8\,$\pm$\,0.6)\% and (6.7\,$\pm$\,0.2)\% (sample 1.) and (39.5\,$\pm$\,0.6)\%, (27.4\,$\pm$\,0.6)\% and (33.1\,$\pm$\,0.6)\% (sample 2.). These values are quoted by the supplier. 

Applying two different anodization voltages (24V and 50V), targets with two different thicknesses were produced. Altogether seven targets were prepared from the two water samples and with the two thicknesses. Intercomparison of the different targets were done by carrying out activation at the same proton energy on targets with different isotopic composition and/or thickness. 

As the determination of the number of target atoms is crucial for the precise cross section measurements, different experimental techniques were used to determine this quantity. First, the Ta:O stoichiometry ratio and the thickness of the oxide layer were measured with Rutherford Backscattering Spectrometry (RBS). The first set of RBS measurements were carried out before the start of the activation experiments at the microbeam setup installed at the 5\,MV Van de Graaff accelerator of Atomki \cite{hus16}. A 1.6\,MeV $\alpha$ beam bombarded the Ta$_2$O$_5$ targets and the scattered particles were detected by two ion implanted Si detectors positioned at 135 and 165 degrees with respect to the beam direction. Exploiting the high lateral resolution of the microbeam setup, spectra were recorded at several different positions on the target surfaces. This test proved that the thickness and stoichiometry of the targets are uniform along the whole surface of the targets. The spectra were analysed using the SIMNRA code \cite{SIMNRA} which provided the areal density of the O atoms as well as the Ta:O ratio.

A second set of RBS measurement was carried out after the activation experiments using a completely independent setup, namely the activation chamber itself (i.e. similar beam size and position to the proton beam used for the activations, see below). At the Tandetron accelerator a 10\,MeV $^{16}$O$^{4+}$ beam bombarded the Ta$_2$O$_5$ targets and a Si detector built into the activation chamber detected the backscattered ions. 

\begin{figure}
\includegraphics[angle=270,width=\columnwidth]{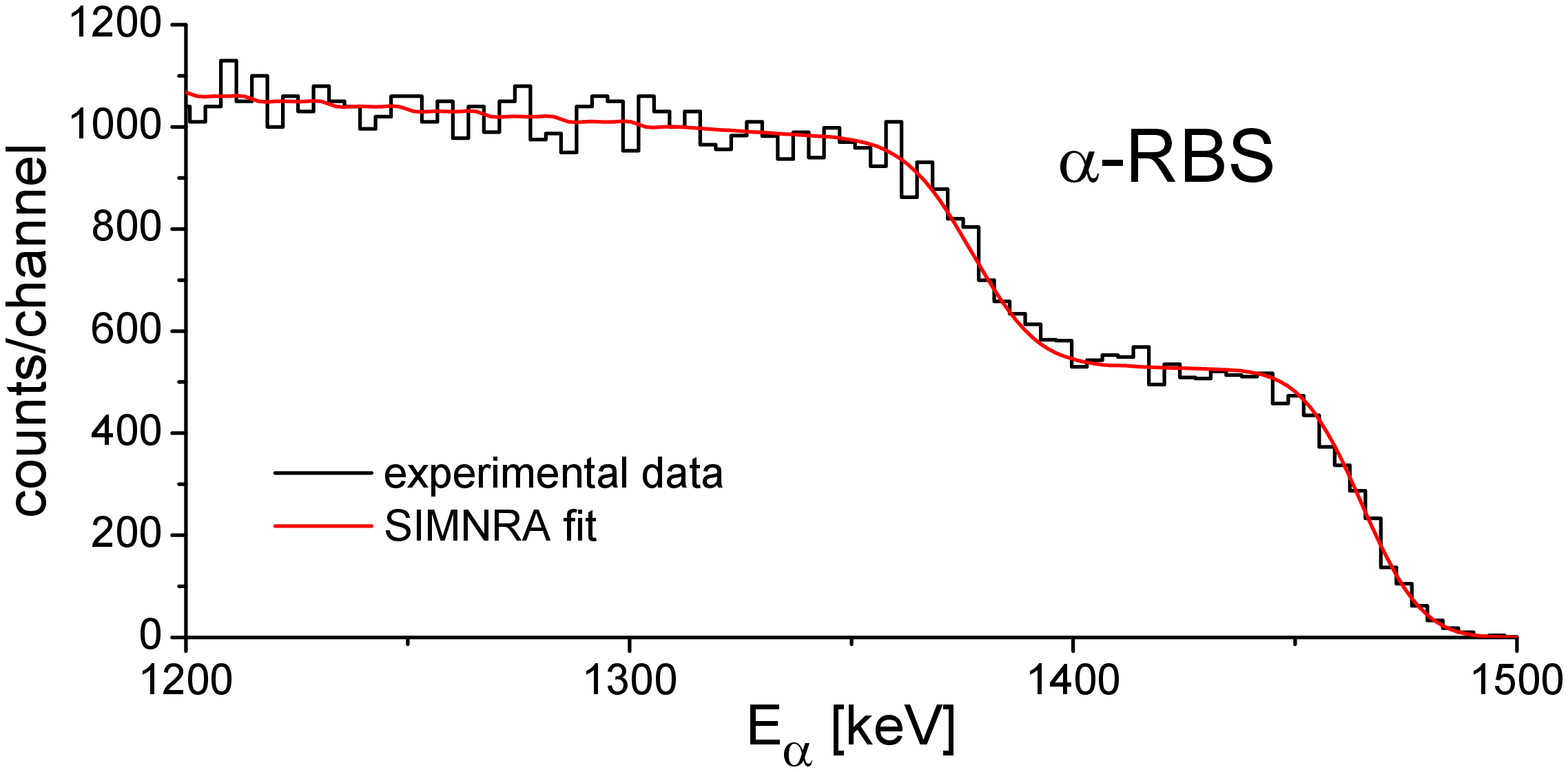}\\
\includegraphics[angle=270,width=\columnwidth]{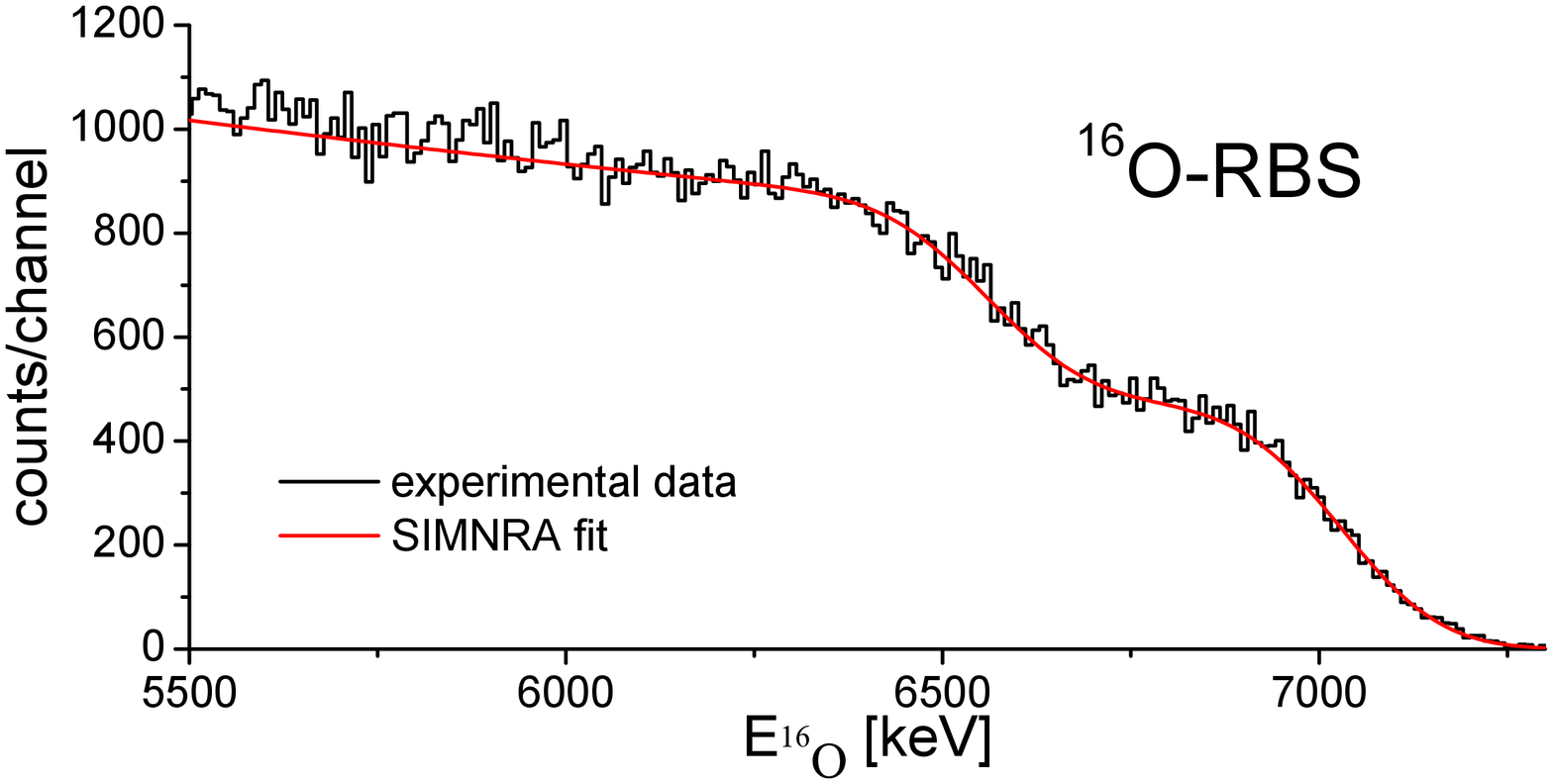}%
\caption{\label{fig:RBS} Relevant parts of the RBS spectra of a Ta$_2$O$_5$ target measured with $\alpha$-beam at the microprobe facility (upper panel) and with $^{16}$O$^{4+}$ beam in the activation chamber at the Tandetron accelerator.}
\end{figure}

Figure \ref{fig:RBS} shows typical spectra of the two RBS measurements. The measured data as well as the fits using the SIMNRA code are shown. The results of the $^{16}$O RBS measurements were in good agreement with the ones obtained with $\alpha$-RBS (see below).
The ratio of the Ta:O atoms was found to be 0.411\,$\pm$\,0.015 in agreement with the stoichiometric value of 0.4.

If the Ta:O ratio is known, a totally independent target thickness value can be obtained by the measurement of the resonance profile on a suitable nuclear resonance. We have investigated the target thicknesses also by this method using both $^{17}$O and $^{18}$O isotope content of the targets. The E$_p$\,=\,1098\,keV and E$_p$\,=\,1925\,keV resonances in the $^{17}$O(p,$\gamma$)$^{18}$F and $^{18}$O(p,$\gamma$)$^{19}$F reactions, respectively, were used to measure the target profiles. A 100\,\% relative efficiency HPGe detector were placed next to the activation chamber at zero degree with its front face about 1\,cm distance from the target. The yield of the strongest transition was used for the measurement of the profiles which was the 937\,keV transition of the first excited state to the ground state in the case of $^{18}$F ($^{17}$O(p,$\gamma$)$^{18}$F reaction) and the 197 keV transition of the second excited state to the ground state in the case of $^{19}$F ($^{18}$O(p,$\gamma$)$^{19}$F reaction). The number of target atoms was obtained from the width of the target profiles using the Ta:O ratio given by the RBS measurements. 

\begin{figure}
\includegraphics[angle=270,width=\columnwidth]{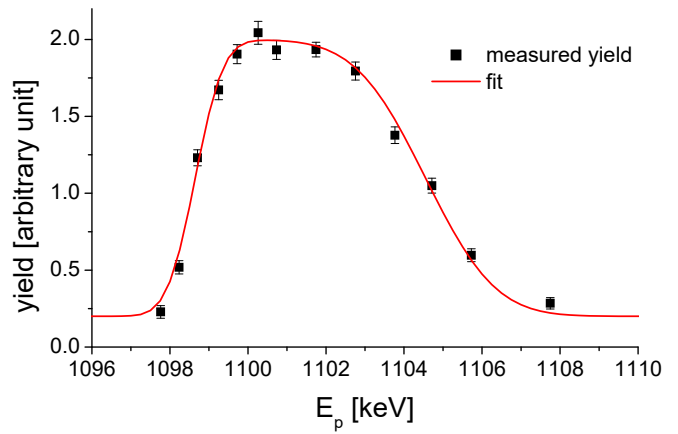}%
\caption{\label{fig:resonance} Measured profile on the E$_p$\,=\,1098\,keV resonance in $^{17}$O(p,$\gamma$)$^{18}$F. The target thickness was obtained from the width of the fitted resonance profile.}
\end{figure}

Figure\,\ref{fig:resonance} shows a typical resonance profile measured with the $^{17}$O(p,$\gamma$)$^{18}$F reaction. The target thickness obtained from the resonance profile measurement using the two reactions gave consistent results. The comparison with the RBS results, however, revealed a roughly 9\,\% systematic difference. The RBS measurements resulted in systematically higher thickness values. Table\,\ref{tab:target} summarizes the thickness results of a given target (prepared with 50\,V anodization voltage) obtained with the four measurements. 

The uncertainties quoted in the table are statistical only stemming form the fit of the RBS spectra and the resonance profiles. Taking into account only these errors, the two methods are in contradiction. If, however, one includes the uncertainty of the stopping power, the results can be considered to be consistent. It is difficult to quantify the uncertainty of stopping power in our experiment as the stopping of protons, $\alpha$-particles and $^{16}$O isotopes should be considered in O and Ta, and the related information \cite{SRIMerror} in the widely used SRIM code indicates uncertainties from about 2\,\% up to 6\,\%. Most likely the deviation of the thickness values has its origin in the uncertainty of the stopping power. Therefore, we have adopted the average of the two methods and assigned a conservative 6\,\% uncertainty to the number of target atoms.

\begin{table}
\caption{\label{tab:target} Results of the various thickness measurements on one of the targets. See text for details.
}
\begin{ruledtabular}
\begin{tabular}{ll}
Method  & No. of O atoms \\
          & 		[10$^{17}$ atoms/cm$^2$]	\\
\hline
$\alpha$-RBS &  5.10\,$\pm$\,0.13 \\
$^{16}$O-RBS &  5.00\,$\pm$\,0.20 \\
$^{17}$O(p,$\gamma$)$^{18}$F resonance &  4.67\,$\pm$\,0.15 \\
$^{18}$O(p,$\gamma$)$^{19}$F resonance &  4.63\,$\pm$\,0.18 \\
\hline
RBS average &  5.07\,$\pm$\,0.11 \\
resonance average &  4.65\,$\pm$\,0.12 \\
\hline
adopted &  4.87\,$\pm$\,0.29 \\
\end{tabular}
\end{ruledtabular}
\end{table}

\subsection{Activations}
\label{subsec:activations}

The activations were carried out at the new Tandetron laboratory of Atomki where a 2\,MV Tandetron accelerator manufactured by High Voltage Engineering Europa B.V. was installed in 2015. The energy calibration of the accelerator was carried out by measuring resonances in the $^{27}$Al(p,$\gamma$)$^{28}$Si reaction and the neutron thresholds in $^{7}$Li(p,n)$^{7}$Be and $^{13}$C(p,n)$^{13}$O reactions \cite{raj16}. 

The Tandetron provided proton beams in the energy range between 500\,keV and 1.8\,MeV and the beam current was limited to about 5\,$\mu$A in order to avoid target degradation. The lifetime of the targets was also increased by using an off-axis target chamber where the beam spot was shifted from the target center by 6\,mm. By rotating the target between the consecutive activations, fresh or not heavily bombarded target spots could be selected. The target chamber was insulated from the rest of the beam line and served as a Faraday cup in order to determine the number of projectiles by charge integration. A secondary electron suppression voltage of -300\,V was applied behind the 4\,mm diameter entrance aperture of the chamber.

Depending on the cross section, the length of the irradiations varied between 15\,minutes and 5 hours. 
Although the beam intensity during the irradiations were typically very stable, in order to follow the possible fluctuations, the beam current was recorded in multichannel scaling mode with one minute time basis. The recorded time dependence of the beam current was then used in the analysis.

\subsection{Measurement of the $^{18}$F decay}
\label{subsec:decay}

After the irradiation the target was removed from the chamber and transported to the counting laboratory where a 100\,\% relative efficiency HPGe detector equipped with full 4\,$\pi$ lead shielding was used to measure the annihilation $\gamma$-radiation of the targets. The $\gamma$-countings started typically 15 minutes after the end of the irradiation and the spectra were recorded in every 10 minutes in order to follow the $^{18}$F decay. 

Since the 511\,keV annihilation radiation is present also in the laboratory background and can come from many possible sources, it is crucial to determine the background. The length of the countings was therefore typically 16 hours. Towards the end of this counting period, the activity of $^{18}$F decayed to a negligible level and therefore the 511\,keV background level could be estimated. This was always found to be consistent with the laboratory background measured without target, indicating that no long-lived positron emitter was created in the targets. 

In some cases excess in the 511\,keV activity was observed at the beginning of the counting period indicating the production of some short-lived positron emitter. From its decay rate it was identified as $^{13}$N produced by the $^{12}$C(p,$\gamma$)$^{13}$N reaction on carbon impurity of the target. This identification was also supported by the fact that such a deviation from the pure $^{18}$F decay was observed mostly around 500\,keV proton energy where the $^{12}$C(p,$\gamma$)$^{13}$N reaction has high cross section due to a broad resonance at about 420\,keV \cite{bur08}.
In such cases roughly the first one hour of the counting was omitted from the analysis. 

\begin{figure}
\includegraphics[angle=270,width=\columnwidth]{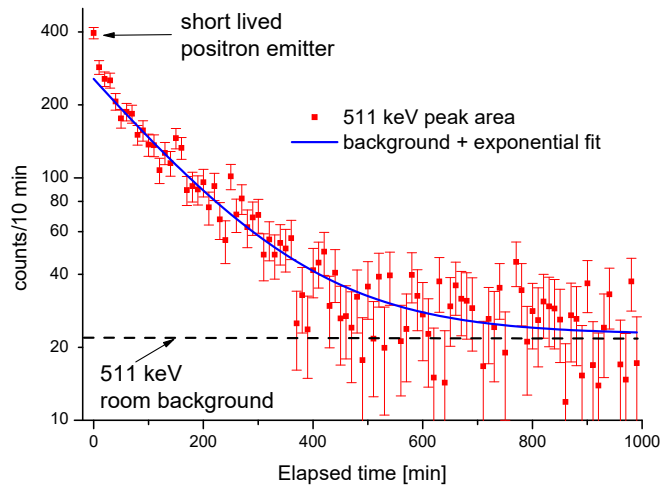}%
\caption{\label{fig:decay} Decay of $^{18}$F in a target irradiated with an 520\,keV proton beam. Taking into account the initial short lived positron emitter and the laboratory background, the decay can be well fitted with the 109.77\,min half-life of $^{18}$F.}
\end{figure}

The decay of the 511\,keV activity could always be fitted well using the literature half-life of $^{18}$F. As an example, Fig.\,\ref{fig:decay} shows the decay curve measured after the irradiation at 520\,keV. The figure indicates the above discussed $^{13}$N contribution, the $^{18}$F decay fitted with the literature half-life and the laboratory background level of the 511\,keV line.

In order to maximize the detection efficiency, the targets were placed in close geometry onto the detector, directly on top of the detector end cap. Since the two annihilation photons are emitted from the source at opposite directions and no other X-ray or $\gamma$-radiation follows the $^{18}$F decay, true coincidence summing effect was not present in this measurement in spite of the close source-to-detector geometry \footnote{The true coincidence of the two annihilation gammas through a Compton scattering process was observed to cause less than 0.5\,\% loss of counts from the 511\,keV peak and was therefore neglected.}. The summing effect, on the other hand, is significant in the case of any multiline calibration source which could be used for the measurement of the absolute detection efficiency. The absolute efficiency was therefore measured in the counting geometry only with single line calibration sources. Calibrated $^{7}$Be, $^{65}$Zn, $^{85}$Sr and $^{137}$Cs sources were used to obtain the efficiency curve of the detector. The $^{85}$Sr source was especially useful for the efficiency determination as it has a single $\gamma$-line at 514\,keV, very close to the relevant 511\,keV. The absolute efficiency was measured with 3\,\% uncertainty which includes also the beam spot size and target positioning effect. 

\section{Experimental results}
\label{sec:results}

The cross section of $^{17}$O(p,$\gamma$)$^{18}$F was measured between proton energies of 500\,keV and 1.8\,MeV. The selection of the actual proton energies was based on the structure of the $^{17}$O(p,$\gamma$)$^{18}$F excitation function. The two broad resonances at 590 and 717\,keV proton energies were measured with fine energy steps. The low energy tails of these resonances have significant contributions to the astrophysical reaction rate of $^{17}$O(p,$\gamma$)$^{18}$F especially at higher temperatures. 

The energy regions where there are no resonances were covered with fewer data points. The aim of these measurements were to fix the absolute value of the direct capture part of the cross section which again has an important contribution to the reaction rate. 

In the studied energy range there are several narrow resonances at proton energies of 517, 673, 741, 826, 926, 1098, 1240, 1270 and 1345 keV. Activation runs at these energies were also carried out with the aim of confirming their existence in the (p,$\gamma$) channel and check their resonance energies. The widths of these resonances, on the other hand, are often comparable with the target thicknesses used in the present work, the determination of the strengths of these resonances was therefore not aimed.

The obtained cross section results are listed in Table\,\ref{tab:results}. The first column shows the energy of the proton beam provided by the Tandetron accelerator. Based on the accelerator calibration, this value is known with a total uncertainty of less than 0.5\,keV. The energy loss of the beam in the target layer is given in the second column. Especially near the resonances the cross section changes significantly in the energy range covered by the target thickness. An effective proton energy was therefore calculated. For this calculation it was supposed that the cross section has a linear energy dependence in the energy range of the target. The slope of the cross section was estimated based on the adjacent experimental data points and on the shape of the excitation function as measured by previous works. The effective energy was then given by the median of the yield curve within the target thickness \cite{lem08}. The uncertainty of the effective energy as listed in the table was estimated based on the slope of the cross section function. Higher energy uncertainties were assigned to the data points near the narrow resonances where the cross section changes strongly within the target thickness.

\begingroup
\begin{table*}
\caption{\label{tab:results} Measured cross section of the $^{17}$O(p,$\gamma$)$^{18}$F reaction in the present work. The quoted cross section uncertainties are statistical only. For the total uncertainty, 7.6\,\% systematic uncertainty must be added quadratically to the relative statistical uncertainties.}
\begin{ruledtabular}
\begin{tabular}{llr@{\hspace{-2mm}}c@{\hspace{-2mm}}lr@{\hspace{-2mm}}c@{\hspace{-2mm}}l||llr@{\hspace{-2mm}}c@{\hspace{-2mm}}lr@{\hspace{-2mm}}c@{\hspace{-2mm}}l}
E$_p$ & Energy loss & \multicolumn{3}{c}{E$_{p\rm ,eff.}$} & \multicolumn{3}{c||}{Cross section}  & E$_p$ & Energy loss & \multicolumn{3}{c}{E$_{p\rm ,eff.}$} & \multicolumn{3}{c}{Cross section}\\
 & in target\footnote{See Section \ref{subsec:target} for information about the target thicknesses} & & & & & & & & in target$^a$ \\
keV & keV & \multicolumn{3}{c}{keV} & \multicolumn{3}{c||}{$\mu$barn} & keV & keV & \multicolumn{3}{c}{keV} & \multicolumn{3}{c}{$\mu$barn}\\
\hline
500.0	&	 4.38 	&	497.8	&	$\pm$	&	 1.4 	&	0.592	&	$\pm$	&	0.021	&	789.7	&	 7.23 	&	786.0	&	$\pm$	&	 2.1 	&	3.30	&	$\pm$	&	0.1	\\
509.8	&	 9.05 	&	505.4	&	$\pm$	&	 2.5 	&	0.587	&	$\pm$	&	0.118	&	819.8	&	 3.41 	&	818.2	&	$\pm$	&	 1.3 	&	2.36	&	$\pm$	&	1.0	\\
514.7	&	 4.32 	&	512.6	&	$\pm$	&	 1.4 	&	0.738	&	$\pm$	&	0.120	&	824.8	&	 3.40 	&	823.2	&	$\pm$	&	 1.3 	&	9.51	&	$\pm$	&	0.3	\\
519.8	&	 4.30 	&	517.7	&	$\pm$	&	 1.4 	&	16.3	&	$\pm$	&	1.8	&	829.7	&	 3.38 	&	828.0	&	$\pm$	&	 1.3 	&	12.2	&	$\pm$	&	0.3	\\
524.7	&	 4.28 	&	522.0	&	$\pm$	&	 2.0 	&	6.13	&	$\pm$	&	0.37	&	834.7	&	 3.37 	&	832.9	&	$\pm$	&	 1.3 	&	2.5	&	$\pm$	&	0.20	\\
529.7	&	 4.26 	&	527.1	&	$\pm$	&	 1.8 	&	1.37	&	$\pm$	&	0.38	&	880.0	&	 3.28 	&	878.4	&	$\pm$	&	 1.2 	&	2.06	&	$\pm$	&	0.13	\\
539.7	&	 4.22 	&	537.7	&	$\pm$	&	 1.4 	&	1.96	&	$\pm$	&	0.23	&	919.8	&	 6.69 	&	916.4	&	$\pm$	&	 1.8 	&	2.18	&	$\pm$	&	0.12	\\
549.7	&	 8.72 	&	545.9	&	$\pm$	&	 2.8 	&	2.93	&	$\pm$	&	0.14	&	924.8	&	 3.20 	&	924.0	&	$\pm$	&	 1.6 	&	1.97	&	$\pm$	&	0.09	\\
559.8	&	 8.64 	&	556.1	&	$\pm$	&	 2.9 	&	4.95	&	$\pm$	&	0.59	&	929.7	&	 6.66 	&	926.4	&	$\pm$	&	 1.9 	&	12.1	&	$\pm$	&	0.2	\\
569.8	&	 4.11 	&	568.0	&	$\pm$	&	 1.6 	&	10.3	&	$\pm$	&	1.0	&	934.7	&	 3.19 	&	933.1	&	$\pm$	&	 1.2 	&	2.65	&	$\pm$	&	0.09	\\
579.7	&	 8.49 	&	576.7	&	$\pm$	&	 3.4 	&	30.5	&	$\pm$	&	0.8	&	999.7	&	 6.43 	&	996.5	&	$\pm$	&	 1.8 	&	2.67	&	$\pm$	&	0.08	\\
584.7	&	 4.06 	&	582.9	&	$\pm$	&	 1.5 	&	67.8	&	$\pm$	&	1.2	&	1089.7	&	 6.17 	&	1086.8	&	$\pm$	&	 1.9 	&	3.55	&	$\pm$	&	0.08	\\
589.7	&	 4.04 	&	587.7	&	$\pm$	&	 1.3 	&	107	&	$\pm$	&	0.8	&	1096.8	&	 2.95 	&	1095.3	&	$\pm$	&	 1.1 	&	11.7	&	$\pm$	&	2.8	\\
599.7	&	 8.35 	&	595.0	&	$\pm$	&	 2.7 	&	70.7	&	$\pm$	&	1.2	&	1101.8	&	 6.14 	&	1098.8	&	$\pm$	&	 1.8 	&	176	&	$\pm$	&	0.8	\\
609.8	&	 8.28 	&	604.5	&	$\pm$	&	 3.3 	&	25.0	&	$\pm$	&	0.4	&	1106.7	&	 6.13 	&	1103.7	&	$\pm$	&	 1.7 	&	35.5	&	$\pm$	&	1.6	\\
619.8	&	 8.21 	&	614.9	&	$\pm$	&	 2.9 	&	10.5	&	$\pm$	&	0.2	&	1111.7	&	 6.12 	&	1108.1	&	$\pm$	&	 2.3 	&	6.03	&	$\pm$	&	0.4	\\
629.7	&	 3.91 	&	627.7	&	$\pm$	&	 1.4 	&	4.14	&	$\pm$	&	0.34	&	1150.0	&	 2.89 	&	1148.6	&	$\pm$	&	 1.1 	&	4.24	&	$\pm$	&	0.1	\\
639.7	&	 3.88 	&	637.8	&	$\pm$	&	 1.3 	&	4.03	&	$\pm$	&	0.21	&	1224.8	&	 5.86 	&	1221.9	&	$\pm$	&	 1.7 	&	5.71	&	$\pm$	&	0.1	\\
649.7	&	 3.84 	&	647.8	&	$\pm$	&	 1.3 	&	3.39	&	$\pm$	&	0.21	&	1239.7	&	 5.82 	&	1237.8	&	$\pm$	&	 2.6 	&	12.2	&	$\pm$	&	1.3	\\
659.8	&	 3.81 	&	657.9	&	$\pm$	&	 1.3 	&	3.78	&	$\pm$	&	0.15	&	1244.7	&	 5.81 	&	1241.9	&	$\pm$	&	 1.7 	&	55.7	&	$\pm$	&	0.7	\\
669.8	&	 7.88 	&	665.8	&	$\pm$	&	 2.1 	&	5.00	&	$\pm$	&	0.51	&	1249.7	&	 2.79 	&	1248.1	&	$\pm$	&	 1.3 	&	19.2	&	$\pm$	&	0.9	\\
672.5	&	 3.78 	&	671.0	&	$\pm$	&	 1.7 	&	138	&	$\pm$	&	1.7	&	1254.7	&	 5.79 	&	1251.4	&	$\pm$	&	 2.1 	&	12.7	&	$\pm$	&	0.4	\\
674.8	&	 3.77 	&	672.8	&	$\pm$	&	 1.3 	&	218	&	$\pm$	&	3.6	&	1259.7	&	 5.78 	&	1256.6	&	$\pm$	&	 1.8 	&	9.34	&	$\pm$	&	0.22	\\
677.3	&	 3.76 	&	674.6	&	$\pm$	&	 1.9 	&	100	&	$\pm$	&	2.3	&	1264.8	&	 5.77 	&	1262.2	&	$\pm$	&	 2.0 	&	8.83	&	$\pm$	&	0.36	\\
679.7	&	 3.76 	&	676.4	&	$\pm$	&	 1.9 	&	16.9	&	$\pm$	&	0.6	&	1274.8	&	 5.75 	&	1271.9	&	$\pm$	&	 1.7 	&	19.5	&	$\pm$	&	0.2	\\
689.7	&	 7.76 	&	685.8	&	$\pm$	&	 2.1 	&	8.72	&	$\pm$	&	0.20	&	1279.8	&	 2.76 	&	1278.3	&	$\pm$	&	 1.1 	&	12.0	&	$\pm$	&	0.3	\\
699.7	&	 3.70 	&	698.0	&	$\pm$	&	 1.4 	&	14.9	&	$\pm$	&	0.93	&	1284.7	&	 2.75 	&	1283.3	&	$\pm$	&	 1.1 	&	7.75	&	$\pm$	&	0.10	\\
704.7	&	 7.68 	&	702.1	&	$\pm$	&	 3.3 	&	26.3	&	$\pm$	&	0.83	&	1299.8	&	 5.70 	&	1296.8	&	$\pm$	&	 1.7 	&	5.19	&	$\pm$	&	0.22	\\
709.7	&	 7.65 	&	707.5	&	$\pm$	&	 3.7 	&	47.0	&	$\pm$	&	1.9	&	1339.7	&	 2.70 	&	1338.5	&	$\pm$	&	 1.1 	&	4.22	&	$\pm$	&	0.14	\\
714.8	&	 3.66 	&	713.1	&	$\pm$	&	 1.4 	&	135	&	$\pm$	&	4.2	&	1345.7	&	 5.62 	&	1343.0	&	$\pm$	&	 1.6 	&	23.1	&	$\pm$	&	0.2	\\
717.8	&	 3.65 	&	716.1	&	$\pm$	&	 1.4 	&	170	&	$\pm$	&	0.8	&	1349.7	&	 2.69 	&	1348.1	&	$\pm$	&	 1.3 	&	5.42	&	$\pm$	&	0.17	\\
719.8	&	 7.59 	&	715.6	&	$\pm$	&	 2.4 	&	178	&	$\pm$	&	1.0	&	1354.7	&	 2.69 	&	1353.3	&	$\pm$	&	 1.1 	&	3.98	&	$\pm$	&	0.16	\\
724.8	&	 7.57 	&	720.3	&	$\pm$	&	 2.7 	&	121	&	$\pm$	&	1.2	&	1359.7	&	 2.68 	&	1358.4	&	$\pm$	&	 1.0 	&	4.33	&	$\pm$	&	0.21	\\
729.7	&	 3.62 	&	727.6	&	$\pm$	&	 1.5 	&	38.6	&	$\pm$	&	1.7	&	1400.0	&	 5.52 	&	1397.2	&	$\pm$	&	 1.6 	&	5.25	&	$\pm$	&	0.12	\\
739.7	&	 7.48 	&	735.3	&	$\pm$	&	 2.7 	&	18.6	&	$\pm$	&	1.0	&	1500.0	&	 2.57 	&	1498.7	&	$\pm$	&	 1.0 	&	5.76	&	$\pm$	&	0.08	\\
744.7	&	 3.58 	&	743.3	&	$\pm$	&	 1.6 	&	12.2	&	$\pm$	&	0.8	&	1600.0	&	 5.18 	&	1597.4	&	$\pm$	&	 1.5 	&	7.94	&	$\pm$	&	0.31	\\
749.7	&	 3.57 	&	747.9	&	$\pm$	&	 1.2 	&	49.6	&	$\pm$	&	0.6	&	1700.0	&	 5.01 	&	1697.5	&	$\pm$	&	 1.5 	&	10.7	&	$\pm$	&	0.3	\\
754.7	&	 3.56 	&	752.9	&	$\pm$	&	 1.2 	&	7.66	&	$\pm$	&	0.2	&	1800.0	&	 2.32 	&	1798.8	&	$\pm$	&	 1.0 	&	11.8	&	$\pm$	&	0.2	\\
759.7	&	 7.38 	&	755.9	&	$\pm$	&	 2.1 	&	7.34	&	$\pm$	&	0.3	\\																
\end{tabular}
\end{ruledtabular}
\end{table*}
\endgroup

In the table only the statistical uncertainty of the cross section values is quoted. This is obtained simply from the peak integration of the 511\,keV $\gamma$-peak and the background subtraction. Typically the statistical uncertainties are between 0.5\,\% and 5\,\%. Higher statistical uncertainties can be found in the case of the lowest cross sections and for those points where based on the literature data higher cross sections were expected at a resonance but the actual resonance was found at slightly shifted energy (see below). 

In order to obtain the total uncertainty of the cross section values, 7.6\,\% systematic uncertainty must be added quadratically to the relative statistical uncertainties. This systematic uncertainty is the quadratic sum of the following components: number of oxygen atoms in the target (6\,\%), $\gamma$-detection efficiency (3\,\%), number of protons hitting the target (3\,\% from charge integration), $^{17}$O enrichment (2\,\%). Uncertainties well below 1\,\% - like the uncertainty of $^{18}$F decay parameters or the measurement of irradiation and counting times - were neglected.

In order to increase the reliability of our experiments, repeated activations were carried out at a few different proton energies using different targets. The results were always in agreement within the statistical uncertainties of the measurements. In table\,\ref{tab:results} either the weighted average of these points are shown or - if targets with different widths were used - the more precise value was adopted.

\section{Discussion}
\label{sec:discussion}

As most of the previous experiments yielded partial cross sections for the various transitions in $^{18}$F measured at a given angle, it is rather difficult to compare the results of the present work with previous experiments. Above 500\,keV proton energy the only total cross section (in the form of an astrophysical S-factor figure) is provided by C. Rolfs \cite{rol73}. Although C. Rolfs studied the reaction in a wide energy range, total S-factor is only provided in the energy regions far from the resonances, i.e. below 450\,keV (outside the energy range of the present work) and above 900\,keV. 

\begin{figure*}
\includegraphics[angle=270,width=0.8\textwidth]{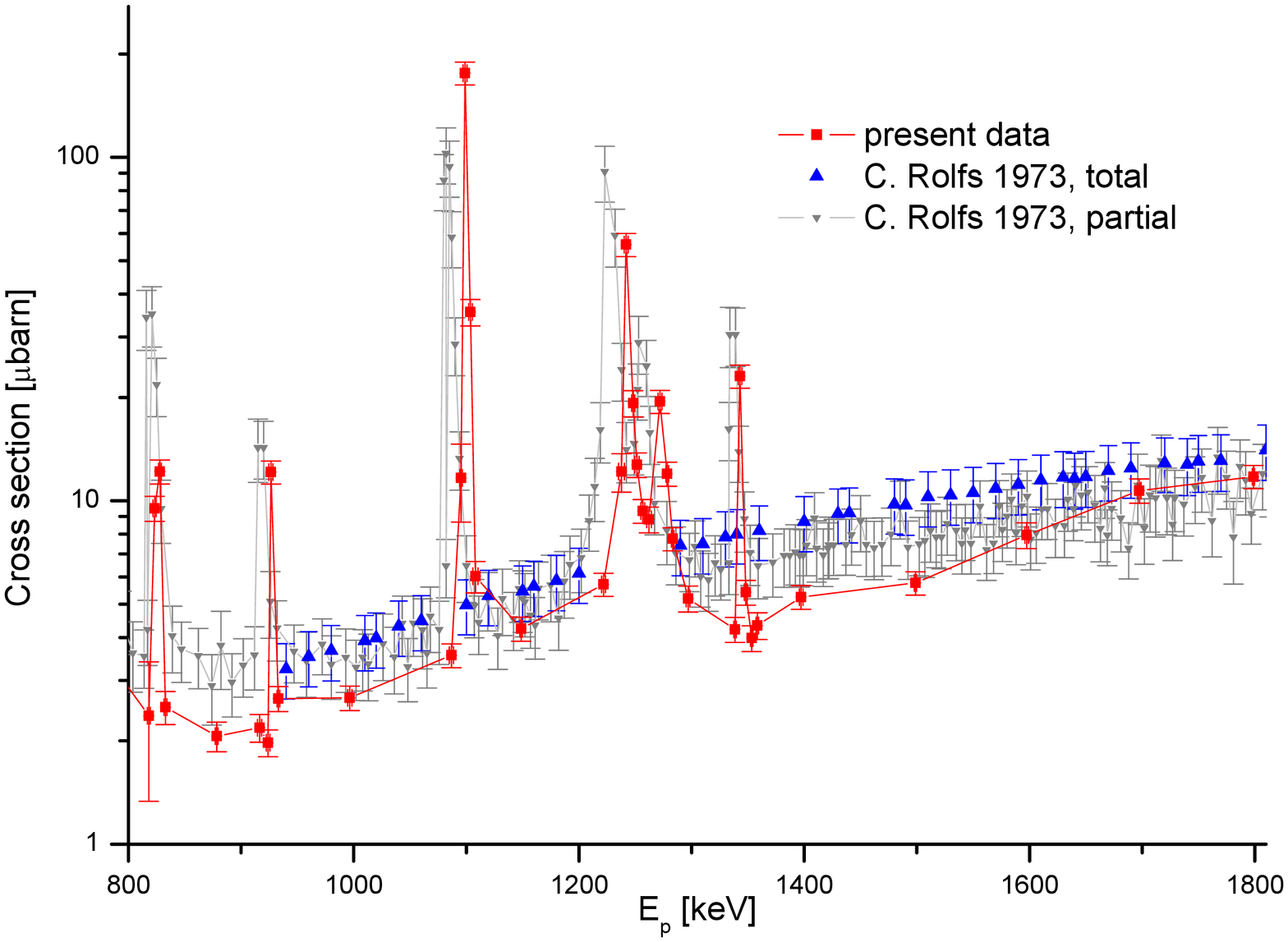}%
\caption{\label{fig:results_highE} Experimental cross section of the $^{17}$O(p,$\gamma$)$^{18}$F reaction in the upper part of the studied energy range. Besides the present work, the total and partial cross sections of C. Rolfs \cite{rol73} are also shown as discussed in the text. The lines through the points are only to guide the eye.}
\end{figure*}

Figure\,\ref{fig:results_highE} shows the total cross section determined in the present work and that of C. Rolfs in this high energy range. The points of C. Rolfs are taken from the EXFOR \cite{EXFOR} database where they were obtained by scanning Fig. 17 of \cite{rol73}. As one can see, the data of C. Rolfs are on average a factor of 1.5 higher than the present data although the agreement becomes somewhat better at the highest energies.

In addition to the total cross section data, the partial cross section involving the first excited state to ground state transition in $^{18}$F measured by C. Rolfs is also included in the figure taken from EXFOR. Obviously, this partial cross section is lower than the total one, but it is included in the figure in order to compare the observed resonances in the high energy region. The existence of the narrow resonances observed by C. Rolfs is confirmed by the present work. There is an apparent energy shift between the two datasets, the resonances in the present work are observed at slightly higher energies than in \cite{rol73} as given in EXFOR. However, the numerical values of the resonance energies, as given in \cite{rol73b} are in reasonably good agreement with the present work. The apparent discrepancy as can be seen in the figure can therefore most likely be attributed to the digitization uncertainty of the low resolution Fig. 11 of \cite{rol73}.

\begin{figure*}
\includegraphics[angle=270,width=0.8\textwidth]{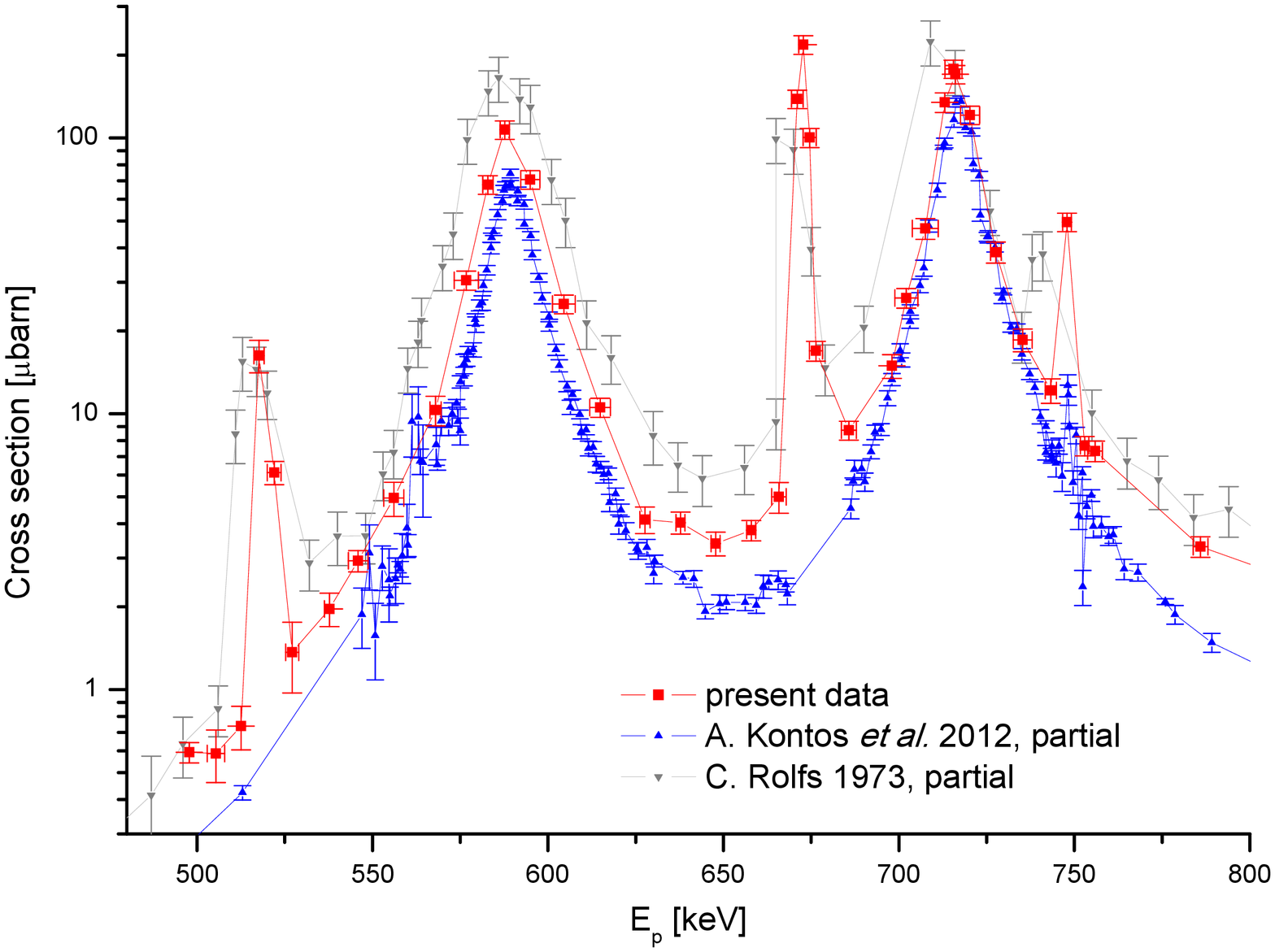}%
\caption{\label{fig:results_lowE} Experimental cross section of the $^{17}$O(p,$\gamma$)$^{18}$F reaction in the lower part of the studied energy range. Besides the present work, the partial cross sections of C. Rolfs \cite{rol73} and A. Kontos \textit{et al.} \cite{kon12} are also shown as discussed in the text. The lines through the points are only to guide the eye.}
\end{figure*}

Between 500\,keV and 900\,keV proton energies no total cross section values are available in literature making the comparison of our data with the existing database even more difficult. Figure\,\ref{fig:results_lowE} shows the measured cross section in this energy range. In order to compare at least the energy dependence of the cross section, besides the present data, partial cross sections measured by C. Rolfs \cite{rol73} and A. Kontos \textit{et al.} \cite{kon12} are also included in the figure. As in Fig.\,\ref{fig:results_highE}, the data of C. Rolfs are taken from its Fig.\,11 as compiled in EXFOR. In the case of A. Kontos \textit{et al.}, capture to the first excited state of $^{18}$F measured at 135 degrees is arbitrarily chosen. The energy dependence of the cross section is very similar in the three datasets. The partial cross section of C. Rolfs exceeds the total cross section measured in the present work. This is similar to the observation on the direct capture cross section at higher energies.

A further comparison with literature data can be made at the lowest studied energy of the present work at E$_p$\,=\,500\,keV. Several recent low energy datasets extend up to this energy and some of them quote total cross section (or S-factor) which can be compared with the present work. Table\,\ref{tab:500keV} lists the experimental (or quasi-experimental, see below) cross section values at 500\,keV proton energy. The following literature data were considered: U. Hager \textit{et al.} \cite{hag12} measured the total cross section with the DRAGON recoil separator at E$_{c.m.}$\,=\,470\,keV corresponding to 497.9\,keV proton energy which matches exactly our lowest energy. The value is taken from table\,VI of \cite{hag12}. In-beam $\gamma$-spectroscopy measurement of J.R. Newton \textit{et al.} \cite{new10} provided total cross section at 500\,keV proton energy which again coincides with our data point taking into account the energy uncertainties. The value is taken from Table\,I of \cite{new10}. A. Kontos \textit{et al.} \cite{kon12} do not provide total cross section data directly at E$_p$\,=\,500\,keV. Measured partial cross section around this energy region, however, is available and based on these data the authors provide total S-factor values in tabular form in their Table\,V. Interpolated value for E$_{c.m.}$\,=\,470\,keV (corresponding to E$_p$\,=\,500\,keV) is put into Table\,\ref{tab:500keV} keeping the 12\,\% relative experimental uncertainty.

\begin{table}
\caption{\label{tab:500keV} Experimental cross section of $^{17}$O(p,$\gamma$)$^{18}$F at E$_p$\,=\,500\,keV from the present and previous works.
}
\begin{ruledtabular}
\begin{tabular}{ll}
Reference  & Cross section at\\
          & 		E$_p$\,=\,500\,keV [nbarn]	\\
\hline
U. Hager \textit{et al.} \cite{hag12}  &  585\,$\pm$\,8$_{\rm stat.}$\,$\pm$\,75$_{\rm syst.}$ \\
J.R. Newton \textit{et al.} \cite{new10} &  488\,$\pm$\,49 \\
A. Kontos \textit{et al.} \cite{kon12}  &  588\,$\pm$\,71\footnote{Not purely experimental value. See text.} \\
\hline
present work &  592\,$\pm$\,21$_{\rm stat.}$\,$\pm$\,45$_{\rm syst.}$ \\
\end{tabular}
\end{ruledtabular}
\end{table}

The result of the present work at E$_p$\,=\,500\,keV is in good agreement with U. Hager \textit{et al.} \cite{hag12} and A. Kontos \textit{et al.} \cite{kon12}. The cross section of J.R. Newton \textit{et al.} \cite{new10}, on the other hand, is almost 20\,\% lower than - and therefore barely consistent with - the other three values.

%
% R-matrix
%

\section{R-matrix analysis}
\label{sec:rmat}

AZURE2, a multichannel and multilevel R-matrix code \cite{azuma10}, was used to simultaneously fit the total cross section, measured by J. R. Newton \textit{et al.} \cite{new10}, U. Hager \textit{et al.} \cite{hag12}, A. Di Leva \textit{et al.} \cite{dil14}, M. Q. Buckner \textit{et al.} \cite{buc15} and by the present work, as well as the primary transitions, measured by A. Kontos \textit{et al.} \cite{kon12}, of the $^{17}$O(p,$\gamma$)$^{18}$F reaction. The fit using these data will be referred to as ``our'' fit in the following.

\begin{figure}
\includegraphics[width=\columnwidth]{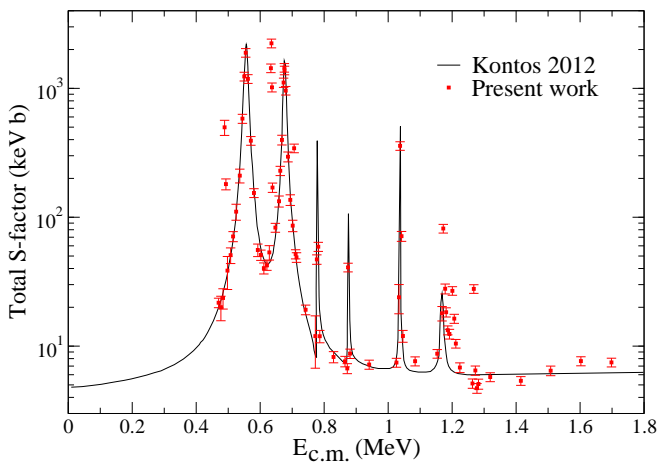}
\caption{\label{fig:Rmat_KontosGyurkyTotalS} Total $S$ factor obtained from an R-matrix fit made by A. Kontos \textit{et al.} \cite{kon12} is compared with our experimental data of the $^{17}$O(p,$\gamma$)$^{18}$F reaction.}
\end{figure}

First, we have compared the total $S$ factor obtained from an R-matrix fit made by A. Kontos \textit{et al.} \cite{kon12} with our experimental data. One can see in Fig.\,\ref{fig:Rmat_KontosGyurkyTotalS} that there is a good agreement between our data and the calculated one of Kontos, although some narrow resonances are omitted from their plot. Values of Kontos are obtained from Fig.\,9 of \cite{kon12} by figure digitization using the software PlotDigitizer 2.6.8 \cite{PlotDig}. 

In our R-matrix fit for the determination of direct capture, because of the nice agreement with the results of Kontos, we used same {\it asymptotic normalization coefficient} (ANC) values and high energy background poles as they used. Table\,\ref{tab:rmatANC} lists these fixed ANC values. In addition, 15 MeV as the excitation energy of the background poles was selected. There are no proton scattering data to provide restrictions for the proton partial widths of the poles, so they were fixed at $\Gamma_p = 6$ MeV, close to the Wigner limit. The R-matrix radius was taken as $r_c = r_0 \times ( A^{1/3}_t+A^{1/3}_p ) = 4.46$ fm, with $r_0 = 1.25$ fm. More details about the selected values are in \cite{kon12}.

\begin{table}
\caption{\label{tab:rmatANC} Fixed ANCs based on \cite{kon12}.}
\begin{ruledtabular}
\begin{tabular}{ c c c }
	Energy (keV) & $\ell$ & ANC (fm$^{-1/2}$)\\[0.05cm]
    \hline\\[-0.2cm]
    ~937 & 0 & 6.1 \\
	~937 & 2 & 1.2 \\
	1121 & 2 & 2.7 \\
	2523 & 0 & 1.4 \\
	3062 & 0 & 4.5 \\
	3062 & 2 & 1.0 \\
	3839 & 0 & 4.6 \\
	3839 & 2 & 0.6 \\
	4115 & 0 & 2.5 \\
	4115 & 2 & 1.0 \\
	4652 & 2 & 1.3 \\
	4964 & 0 & 3.2 \\
	4964 & 2 & 0.7 \\
  \end{tabular}
\end{ruledtabular}
\end{table}

Our R-matrix analysis used the data set of table\,\ref{tab:results}. 
As the quoted effective energies were used, no target effect was taken into account. No normalization of datasets was applied and for physical parameters the Brune parameterization \cite{bru02} was used. The full parameter list of our R-matrix fit is provided as Supplemental Material \cite{supmat}. It contains the used datasets and the AZURE2 input file with all parameters.

The estimated dependence of the R-matrix extrapolation on the choice of the channel radius, the position of the background poles and ANC values are $\sim$\,4\,\%, $\sim$\,7\,\% and $\sim$\,15\,\%, respectively. These values are estimated from the manual variation of the above parameters around their fixed values. The uncertainty of the extrapolation of the total $S$ factor to zero energy is $\sim$\,20\,\%.

Table\,\ref{tab:rmatS0} lists the calculated contributions of all the measured transitions to the total $S$ factor at zero energy. The second column is calculated by Kontos \textit{et al.} \cite{kon12}, the third one by Di Leva \textit{et al.} \cite{dil14} and the last one comes from our fit. The error of our data is $\sim$\,15\,\% because of the uncertainty of the choice of ANC. The uncertainties are statistical only. The total $S$(0) value of Kontos and Di Leva are $5.4 \pm \text{(th.)}1.0 \pm \text{(exp.)}0.6$ keV\,b and $5.0 \pm 0.3$ keV\,b, respectively. Our total $S$ factor value at zero energy is $4.7 \pm 1.0$ keV\,b where the error is statistical only.

\begin{table*}
\caption{\label{tab:rmatS0} Calculated $S$(0) values for each $\gamma$-ray transitions measured by Kontos \textit{et al.} \cite{kon12}.}
\begin{ruledtabular}
\begin{tabular}{ c c c c}
	Transition (keV) & $S$(0)$^{\text{Kontos \cite{kon12}}}$ (keV\,b) & $S$(0)$^{\text{Di Leva \cite{dil14}}}$ (keV\,b) & $S$(0)$^{\text{Present}}$ (keV\,b)\footnote{The $S$(0) of the present work was obtained by using for the fit simultaneously the partial cross sections from the literature (see text) and the total cross section presented in this paper.}\\[0.05cm]
    \hline\\[-0.2cm]
    R/DC $\rightarrow 937~\,$ & $1.7\pm 0.3$ & $1.48\pm 0.08$ & $1.73 \pm 0.26$\\
	R/DC $\rightarrow 1121$ & $0.66\pm 0.13$ & $0.47\pm 0.05$ & $0.65 \pm 0.10$ \\
	R/DC $\rightarrow 1700$ & $0.013\pm 0.002$ &  & $0.013 \pm 0.002$ \\
	R/DC $\rightarrow 2523$ & $0.17\pm 0.03$ & $0.12\pm 0.03$ & $0.15 \pm 0.02$ \\
	R/DC $\rightarrow 3062$ & $0.66\pm 0.1~\,$ & $0.59\pm 0.03$ & $0.45 \pm 0.07$ \\
	R/DC $\rightarrow 3791$ & $0.032\pm 0.005$ & $0.20\pm 0.05$ & $0.030 \pm 0.005$ \\
	R/DC $\rightarrow 3839$ & $0.93\pm 0.14$ & $0.92\pm 0.04$ & $0.66 \pm 0.10$ \\
	R/DC $\rightarrow 4115$ & $0.55\pm 0.08$ & $0.50\pm 0.03$ & $0.51 \pm 0.08$ \\
	R/DC $\rightarrow 4652$ & $0.21\pm 0.03$ & $0.10\pm 0.03$ & $0.19 \pm 0.03$ \\
	R/DC $\rightarrow 4964$ & $0.49\pm 0.07$ & $0.43\pm 0.03$ & $0.35 \pm 0.05$ \\
  \end{tabular}
\end{ruledtabular}
\end{table*}

\begin{figure}
\includegraphics[width=\columnwidth]{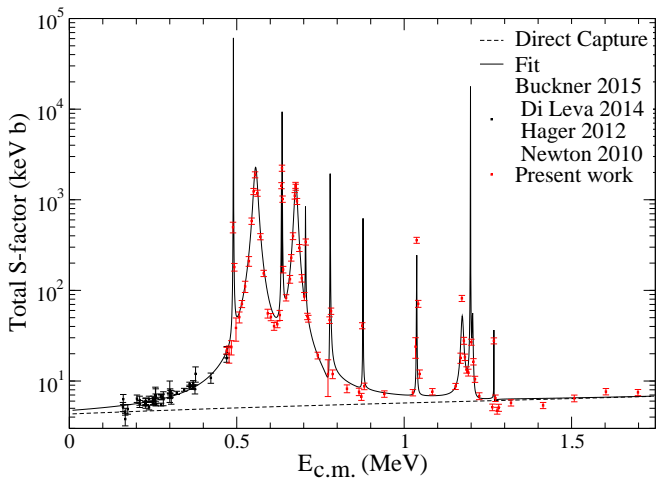}
\caption{\label{fig:Rmat_FitGyurkyTotalS} Total $S$ factor obtained from our R-matrix fit (continuous line) is compared with experimental data of the $^{17}$O(p,$\gamma$)$^{18}$F reaction. The dashed line is the contribution of the direct capture to the total $S$ factor (background poles included).}
\end{figure}

\begin{figure}
\includegraphics[width=\columnwidth]{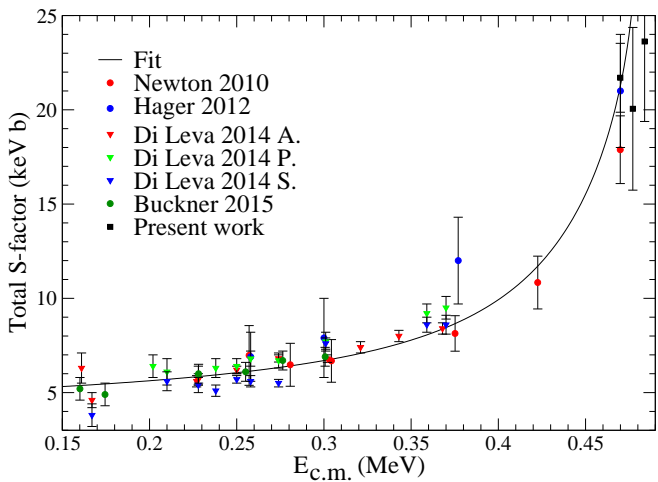}
\caption{\label{fig:Rmat_FitS0} Low energy total $S$ factor obtained from our R-matrix fit is compared with experimental data of the $^{17}$O(p,$\gamma$)$^{18}$F reaction. (A.: activation; P.: primary transitions; S.: secondary transitions)}
\end{figure}

Fig.\,\ref{fig:Rmat_FitGyurkyTotalS} shows the total $S$ factor obtained from our R-matrix fit (continuous line) as well as experimental datasets of J. R. Newton \textit{et al.} \cite{new10}, U. Hager \textit{et al.} \cite{hag12}, A. Di Leva \textit{et al.} \cite{dil14}, M. Q. Buckner \textit{et al.} \cite{buc15} and present work. Narrow resonances are also included. The $\chi^2$ value of our dataset is $7.3$ without any normalization of datasets. The contribution of the direct capture to the total $S$ factor at zero energy in our fit is $S_{\text{DC}}=4.3\pm1.0$ keV\,b, where the uncertainty is statistical only. Fig.\,\ref{fig:Rmat_FitS0} shows the low energy total $S$ factor obtained from our R-matrix fit with the above experimental datasets.

\section{Summary and conclusions}
\label{sec:summary}

In the present work the total cross section of the $^{17}$O(p,$\gamma$)$^{18}$F reaction was measured with the activation method in a wide energy range for the first time with a total uncertainty of about 10\,\%. Since this method provides cross sections which are in several aspects independent from the ones obtained with in-beam $\gamma$-spectroscopy and some systematic errors are not present, our results can be used to check the validity of previous data. 

In general, our data is in good qualitative agreement with the structure of the excitation function of previous works. The possibility of the comparison of total cross sections is limited owing to the scarcity of total cross section data in the literature in the studied energy range. At energies above 900\,keV our results are on average a factor of 1.5 lower than that of C. Rolfs \cite{rol73}. Similar deviation is found at lower energies where the partial cross sections of C. Rolfs exceed substantially our total cross section. Too high values of C. Rolfs were also pointed out earlier by A. Kontos \textit{et al.} \cite{kon12} in the case of individual transitions. This observation is confirmed by the present work.

A direct comparison of our cross section data with the literature was carried out also at a single proton energy of 500\,keV. It is found that our value agrees well with that of U. Hager \textit{et al.} \cite{hag12} and A. Kontos \textit{et al.} \cite{kon12}, while the result of J.R. Newton \textit{et al.} \cite{new10} is about 20\,\% (two standard deviations) lower.

An  R-matrix analysis with the AZURE2 code was performed to check the conformity of our measured total cross section dataset and to extrapolate the astrophysical $S$ factor to lower energies. In this analysis all primary transitions observed from Ref. \cite{kon12} were simultaneously fitted with some total cross section datasets, included the present one. The resulting total $S$ factor is in good agreement with previous measurements and calculations within the experimental uncertainties.

Our total cross section data can be used to constrain any future theoretical description of the $^{17}$O(p,$\gamma$)$^{18}$F reaction. 

\begin{acknowledgments}
We thank A. Formicola and the LUNA collaboration for giving us access to the Ta$_2$O$_5$ target preparation device at LNGS, Italy. We also thank R. J. deBoer for all of his helpful advices regarding the use of the R-matrix code, AZURE2. This work was supported by the SROP-4.2.2.B-15/1/KONV-2015-0001 project, by the European Union, co-financed by the European Social Fund and by OTKA grants No. K108459, K120666 and K112962.%
\end{acknowledgments}

\end{document}